\documentstyle[pra,aps,preprint]{revtex}
\begin{document}

\draft
\title{Composite Fermion Pairing in Bilayer Quantum Hall Systems}
\author{Takao Morinari}
\address{Department of Applied Physics, University of Tokyo,
Bunkyo-ku, Tokyo 113, Japan}

\date{\today}
\maketitle
\begin{abstract}
We derive the effective Hamiltonian for the composite fermion
in double-layer quantum Hall systems with inter-layer tunneling
at total Landau-level filling factor $\nu=1/m$, where $m$ is an integer.
We find that the ground state is the triplet {\it p}-wave BCS pairing state
of the composite fermions.
At $\nu=1/2$, the ground state of the system evolves 
from the Halperin $(3,3,1)$-state toward the {\it Pfaffian}-state 
with increasing the tunneling amplitude. 
On the other hand, at $\nu=1$, the pairing state is uniquely determined 
independent of tunneling amplitude.
\end{abstract}

The fractional quantum Hall effect was discovered \cite{tsui} 
in the two-dimensional electron systems under strong magnetic field
perpendicular to the plane.
The theoretical understanding of this effect was achieved by the Laughlin
wave function \cite{laughlin} at the Landau level filling factor 
$\nu=1/m$, where $m$ is an odd integer.
At other filling factors, 
$\nu=m/(mp+1)$ ($m$ is an integer, $p$ is an even integer), 
the composite fermion (CF) picture \cite{jain} 
is one of the most appealing theoretical frameworks.
The fractional quantum Hall effect of electrons is mapped
into the {\it integral} quantum Hall effect of the CFs.

The CF picture is also useful to understand the fractional quantum Hall effect 
with internal degrees of freedom.
Several years ago, the quantum Hall effect was observed 
in double-layer systems \cite{exp}.
Many numerical works have shown \cite{numerical}
that in the absence of inter-layer tunneling the systems are well described 
by the Halperin $(m,m,n)$-state \cite{halperin}, 
which is ageneralization of the Laughlin wave function 
to the multi-component systems.
Especially, at total filling factor $\nu=1/2$, the system is described by the
$(3,3,1)$-state.
On the other hand, at this filling factor, 
the {\it Pfaffian}-state \cite{moore_read} 
was proposed in the strong tunneling limit \cite{greiter}.
Recently, Ho showed that both the $(3,3,1)$-state
and the {\it Pfaffian}-state belong to the family of triplet {\it p}-wave 
pairing states of the CF by the analysis of wave functions \cite{ho}. 
The CF pairing state is also studied by Bonesteel et al. \cite{bonesteel} 
at $\nu=1/m$. They discussed the CF pairing instability induced by 
the Chern-Simons gauge field fluctuation.
In these works, however, an origin of the pairing potential is unclear and
the inter-layer tunneling effect is discussed only at the qualitative level.

In this paper, we make clear the origin of the pairing 
interaction and derive the effective Hamiltonian in double-layer quantum Hall
systems at $\nu=1/m$ ($m$ is an integer) with the inter-layer tunneling.
Deriving the gap equation and analyzing it we show that the ground state
of the system is the {\it triplet} {\it p}-wave pairing state.
As far as we know, this is the first time 
that the {\it p}-wave BCS pairing state
is established with revealing the origin of pairing potential.
Including the tunneling effect, 
we show quantitatively that at $\nu=1/2$ 
the system evolves from the $(3,3,1)$-state 
toward the {\it Pfaffian}-state.
Furthermore, at $\nu=1$, we show that the system is also described by 
the {\it triplet} {\it p}-wave pairing state,
which corresponds to the $(1,1,1)$-state in the absence of tunneling.
Finally we mention the CF pairing states at the other filling fractions.

To begin with, we discuss the Coulomb interaction effect.
As mentioned above, many numerical calculations support \cite{numerical} 
that the ground state of the bilayer quantum Hall systems is 
well described by the Halperin $(m,m,n)$ wave function \cite{halperin}
in the absence of inter-layer tunneling.
The index $m$($n$) corresponds to the relative angular momentum of the
pair of electrons in the same(opposite) layer.
At given filling factor, 
the indices $m$ and $n$ are fixed by a consideration of the two-body problem
by the following reason.
The Halperin $(m,m,n)$-wave functions are the Jastrow-type wave functions.
As in the $^4$He case, this Jastrow-factors are factorized into two parts:
a short-range component and a long-range component \cite{mahan}. 
The former is determined by the two-body problem and the latter 
by the phonon effect. However, the phonon modes are pushed up to the high
energy modes because of the incompressible nature of quantum Hall systems.
Hence the consideration of the two-body problem completely 
determines $m$ and $n$.
The Coulomb interaction in double-layer quantum Hall systems is given by
\begin{equation}
V_{\alpha \beta}({\bf r}) = 
\frac{e^2}{\epsilon \sqrt{r^2 + (1-\delta_{\alpha \beta}) d^2}},
\end{equation}
where indices $\alpha,\beta = \uparrow, \downarrow$ label the layers,
$\epsilon$ is the dielectric constant and $d$ is the inter-layer separation.
To analyze the two-body problem we take the lowest Landau level wave function
as the basis and estimate the Coulomb energy $E^{(2)}_C$ 
by first order perturbation under the condition of $m+n=2/\nu$.
Numerical estimation of $E^{(2)}_C$ shows that 
the pair $(m,n)$ giving the lowest $E^{(2)}_C$ is $(2/\nu,0)$ 
for $d \gg \ell_B$ and with decreasing $d$, 
changes as $(2/\nu,0) \rightarrow (2/\nu-1,1) \rightarrow \cdots
\rightarrow (1/\nu,1/\nu)$.
For instance, at $\nu=1$, 
we find that $(m,n)=(2,0)$ for $d>d_c$ and
$(m,n)=(1,1)$ for $0 \le d<d_c$ where $d_c/\ell_B \sim 1.4$ 
($\ell_B$ the magnetic length).

In the following discussion we restrict ourselves to the cases
$\nu=1/m$, where $m$ is an integer.
At this filling fraction, the quantum Hall effect occurs 
when $(m,n)=(p,q)$ where both $p$ and $q$ are
{\it odd} integers. To take into account the correlation effect,
we map the interacting electron gas into a system of composite particles 
by attaching magnetic fluxes to the particles.
Let $\phi_1$($\phi_2$) denote the number of fluxes
seen by other particles in the same(opposite)-layer.
The case $(\phi_1,\phi_2)=(p,q)$ reduces to the composite boson theory
 \cite{zhang}. 
This composite boson theory is partially successful 
in describing the double-layer quantum Hall systems \cite{boson}.
However, in this theory the effects of tunneling and inter-layer separation
are unclear. 
On the other hand, to describe the quantum Hall states in double-layer systems 
by the CFs \cite{lopez_fradkin},
there are two choices : $(\phi_1,\phi_2)=(p+1,q-1)$ 
or $(p-1,q+1)$.
Though the former is possible to describe the quantum Hall systems,
we consider the latter by the following reason.
As was pointed out by Haldane and Rezayi \cite{haldane_rezayi} 
and Ho \cite{ho},
the {\it p}-wave fermion BCS \cite{bcs} 
pairing state is equivalent to the $(1,1,-1)$-state.
Therefore, the CF with $(\phi_1,\phi_2)$ $+$ 
{\it p}-wave pairing state is corresponds to 
the $(\phi_1+1,\phi_1+1,\phi_2-1)$-state 
in the absence of inter-layer tunneling.

With this appropriate choice of flux numbers $(\phi_1,\phi_2)$ for the CF,
we shall consider how the remaining correlation effect changes the CF state.
To see this, we map the electron system into the CF system.
The second quantized Hamiltonian 
for the double-layer two-dimensional electron system
in the presence of an external uniform magnetic field $B$ perpendicular to 
the layer is given by $H=H_0+V_C$,
\begin{equation}
H_0  = \sum_{\alpha} \int d^2 {\bf r} \frac{1}{2M} 
           \psi^{\dagger}_{\alpha} ({\bf r}) (-i \nabla + {\bf A})^2 
           \psi_{\alpha} ({\bf r}), 
\end{equation}
\begin{equation}
V_C  =  \int d^2 {\bf r}_1 \int d^2 {\bf r}_2 
        V_{\alpha \beta} ({\bf r}_1 - {\bf r}_2)
        \delta \rho_{\alpha} ({\bf r}_1)
        \delta \rho_{\beta} ({\bf r}_2), 
\end{equation}
($\hbar=c=e=1$)
where $\delta \rho_{\alpha}({\bf r})= \psi^{\dagger}_{\alpha}({\bf r}) 
\psi_{\alpha}({\bf r}) - {\bar \rho_{\alpha}}$ 
with ${\bar \rho_{\alpha}}$ the average
particle density in the layer $\alpha$.
In this paper we concentrate on the case of 
$\bar{\rho_{\uparrow}} = \bar{\rho_{\downarrow}}$.

We introduce generalized CF field operators \cite{rajaraman} by
\begin{equation}
\phi_{\alpha} ({\bf r}) = e^{-J_{\alpha} ({\bf r})} \psi_{\alpha} ({\bf r}),
\label{non_unitary1}
\end{equation}
\begin{equation}
\pi_{\alpha} ({\bf r}) = \psi^{\dagger}_{\alpha} ({\bf r}) 
                         e^{J_{\alpha} ({\bf r})},
\label{non_unitary2}
\end{equation}
where
\begin{equation}
J_{\alpha} ({\bf r}) = \sum_{\beta} K_{\alpha \beta} 
                       \int d^2 {\bf r}^{\prime} 
                        \rho_{\beta} ({\bf r}^{\prime}) \log (z-z^{\prime})
                        -\frac{1}{4{\ell_B}^2} |z|^2,
\label{non_unitary3}
\end{equation}
with $z=x+iy$.
The matrix $K$ is given by
\begin{equation}
K = \left( \begin{array}{rr}
            \phi_1 & \phi_2 \\
            \phi_2 & \phi_1
           \end{array} \right).
\end{equation}
Because $\phi_1$ and $\phi_2$ are even integers, 
the operators $\phi_{\alpha} ({\bf r})$ and $\pi_{\alpha} ({\bf r})$
satisfy the fermion anticommutation relations;
\begin{equation}
\{\phi_{\alpha} ({\bf r}_1),\pi_{\beta} ({\bf r}_2)\} 
        = \delta_{\alpha \beta} \delta^2 ({\bf r}_1 - {\bf r}_2),
\end{equation}
\begin{equation}
\{\phi_{\alpha} ({\bf r}_1),\phi_{\beta} ({\bf r}_2)\} = 0,
\{\pi_{\alpha} ({\bf r}_1),\pi_{\beta} ({\bf r}_2)\} = 0.
\end{equation}

In terms of the CF operators, $H_0$ is described by 
\begin{equation}
H_0^{CF} = \sum_{\alpha} \int d^2 {\bf r} \frac{1}{2M}
           \pi_{\alpha} ({\bf r})
       ( -i \nabla + \delta {\bf a}_{\alpha} 
  + i{\hat z} \times \delta{\bf a}_{\alpha})^2 
           \phi_{\alpha} ({\bf r}),
\label{cf_hamiltonian}
\end{equation}
where 
\begin{equation}
\delta {\bf a}_{\alpha} ({\bf r}) 
 =  \sum_{\beta} K_{\alpha \beta}
      \int d^2 {\bf r}^{\prime} 
    \left( \pi_{\alpha}({\bf r}^{\prime}) \phi_{\alpha} ({\bf r}^{\prime}) 
              - \bar{\rho_{\alpha}} \right)
      \nabla {\rm Im} \log (z-z^{\prime}).
\label{chern_simons}
\end{equation}
Substituting (\ref{chern_simons}) into (\ref{cf_hamiltonian}) and
including the inter-layer tunneling, we obtain
\begin{equation}
H_0^{CF} = \sum_{{\bf k} \alpha \beta} \left( \xi^k \right)_{\alpha \beta}
\pi_{\alpha} ({\bf k}) \phi_{\beta} ({\bf k}) 
+ V_1 + V_2,
\label{cf_hamiltonian2}
\end{equation}
where $\left( \xi^k \right)_{\alpha \beta} = \xi_k \delta_{\alpha \beta}
-t(1-\delta_{\alpha \beta})$($\xi_k=k^2/2M - \mu$, 
$t$ is the tunneling amplitude), and
\begin{equation}
V_1 = \frac{2 \pi i}{M \Omega} \sum_{{\bf k}_1, {\bf k}_2, {\bf q}\neq 0}
  \sum_{\alpha \beta} K_{\alpha \beta} 
   \frac{{\bf k}_1 \times {\bf q}}{q^2} 
   \pi_{\alpha}({\bf k}_1 + {\bf q}) \pi_{\beta} ({\bf k}_2)
   \phi_{\beta}({\bf k}_2 + {\bf q}) \phi_{\alpha} ({\bf k}_1),
\end{equation}
\begin{equation}
V_2 = -\frac{ \pi }{M \Omega} \sum_{{\bf k}_1, {\bf k}_2, {\bf q}\neq 0}
  \sum_{\alpha \beta} K_{\alpha \beta} 
   \frac{q^2 + 2{\bf k}_1 \cdot {\bf q}}{q^2} 
   \pi_{\alpha}({\bf k}_1 + {\bf q}) \pi_{\beta} ({\bf k}_2)
   \phi_{\beta}({\bf k}_2 + {\bf q}) \phi_{\alpha} ({\bf k}_1)
\end{equation}
with $\Omega$ the area of the system.
Here we measure the kinetic energy from the chemical potential $\mu$.
To derive the pairing interaction, we pick up the terms with 
${\bf k}_1 + {\bf k}_2 + {\bf q}=0$ in the ${\bf q}$ summation 
of $V_1$ and $V_2$. 
Then $V_2$ manifestly breaks the 
time reversal symmetry, which shows that 
this potential is an artifact of {\it non-unitary}
transformations (\ref{non_unitary1}), (\ref{non_unitary2}) and
(\ref{non_unitary3}). The pairing potential is derived from $V_1$.
If we drop the $V_2$ term, the analysis is the same as 
in the case of {\it unitary} 
transformation when we replace $\pi$ by $\phi^{\dagger}$.
With this replacement, the final form of the effective Hamiltonian
of the system is given by 
\begin{equation}
H^{eff} = H^{eff}_0 +V_C,
\end{equation}
\begin{equation}
H^{eff}_0 = \sum_{{\bf k} \alpha \beta} \left( \xi^k \right)_{\alpha \beta}
\phi^{\dagger}_{\alpha} ({\bf k}) \phi_{\beta} ({\bf k}) 
+ \frac{1}{2\Omega} \sum_{{\bf k}_1 \neq {\bf k}_2 }
  \sum_{\alpha \beta} K_{\alpha \beta} V_{{\bf k}_1 {\bf k}_2}
   \phi^{\dagger}_{\alpha}({\bf k}_1) \phi^{\dagger}_{\beta} (-{\bf k}_1)
   \phi_{\beta}(-{\bf k}_2) \phi_{\alpha} ({\bf k}_2),
\label{pairing_potential}
\end{equation}
where $V_{{\bf k}_1 {\bf k}_2} = -(4 \pi i /M)
{\bf k}_1 \times {\bf k}_2/|{\bf k}_1-{\bf k}_2|^2$ \cite{spinless}.
To make the pairing state clear, we neglect 
the Coulomb interaction term for a while.
Later we discuss the effect of the Coulomb interaction.

In the same way as the analysis of pairing state of $^3$He \cite{bw},
we derive the equation for the gap 
$\left( \Delta^{\bf k} \right)_{\alpha \beta}$ \cite{meanfield}.
For the triplet pairing case, since there are many types of solutions,
we introduce two assumptions to simplify the discussions.
First, since two layers are symmetric, we assume 
$\left( \Delta^{\bf k} \right)_{\uparrow \uparrow}
=\left( \Delta^{\bf k} \right)_{\downarrow \downarrow}$.
In this case, the gap equation has the block diagonalized form and is given by
\begin{equation}
\left( \Delta^{\bf k} \right)_{\alpha \beta}
= - \frac{1}{2 \Omega} K_{\alpha \beta} \sum_{ {\bf k}^{\prime}(\neq {\bf k}) }
V_{ {\bf k} {\bf k}^{\prime} }
\left[ \Delta^{ {\bf k}^{\prime} } \left( E^{ {\bf k}^{\prime} } \right)^{-1}
\tanh \left( \frac{ E^{ {\bf k}^{\prime} }}{2 k_B T} \right) 
\right]_{\alpha \beta},
\label{block_gap}
\end{equation}
where $E^{\bf k}$ and $\Delta^{\bf k}$ are $2\times 2$-matrix 
and $(E^{\bf k})^2=(\xi^k)^2 + (\Delta^{\bf k})^{\dagger} \Delta^{\bf k}$.
Secondly, we assume that each component of $\Delta^{\bf k}$ has
the same angular $\theta_{\bf k}$ dependence ($\theta_{\bf k}$ denotes
the direction of ${\bf k}$).
The second assumption is introduced for convenience and 
the result about the pairing state of the ground state 
given below holds without it.
From these assumptions, the gap $\Delta^{\bf k}$ is given by
\begin{equation}
\Delta^{\bf k} = \Delta_{\bf k} 
  \left( \begin{array}{rr}
         a & b \\
         b & a \\
         \end{array} \right),
\label{gap_form}
\end{equation}
where $a$ and $b$ are complex numbers satisfying $|a|^2+|b|^2=1$ and
$\Delta_{\bf k} = e^{-i \ell \theta_k} |\Delta_k|$ for
the $\ell$-wave pairing. The gap equation for the ground state is given by,
\begin{equation}
a \Delta_{\bf k} = - \frac{\phi_1}{4} 
\left[ (a+b) P^+_{\bf k} + (a-b) P^-_{\bf k}) 
\right],
\label{gap_equation1}
\end{equation}
\begin{equation}
b \Delta_{\bf k} = - \frac{\phi_2}{4} 
\left[ (a+b) P^+_{\bf k} - (a-b) P^-_{\bf k}
\right].
\label{gap_equation2}
\end{equation}
Here $P^{\pm}_{\bf k}$ is given by
\begin{eqnarray}
P^{\pm}_{\bf k}
& \equiv &
 \frac{1}{\Omega} \sum_{{\bf k}^{\prime}(\neq {\bf k})} 
V_{{\bf k} {\bf k}^{\prime}}
 \frac{\Delta_{{\bf k}^{\prime}}}{E^{\pm}_{k^{\prime}}} \nonumber \\
& = & 
\frac{1}{M} e^{-i \ell \theta_k} 
\left( \int_0^{k} dk^{\prime} + \int_k^{\infty} dk^{\prime} \right)
\frac{k^{\prime} |\Delta_{k^{\prime}}|}{E^{\pm}_{k^{\prime}}}
I_{\ell} (k,k^{\prime}),
\label{eq_P}
\end{eqnarray}
where $E^{\pm}_k=\sqrt{(\xi_k -(\pm t))^2 + |a \pm b|^2 |\Delta_k |^2}$,
\begin{equation}
I_{\ell} (k,k^{\prime}) = \int_0^{2\pi} \frac{d \theta}{2\pi i}
\frac{e^{-i \ell \theta} \sin \theta}{(k^2+{k^{\prime}}^2)/2 k k^{\prime}
- \cos \theta}.
\end{equation}
In the case of $\ell > 0$,
\begin{equation}
I_{\ell}(k,k^{\prime}) = \left\{ \begin{array}{ll}
-(k^{\prime}/k)^{\ell} & \mbox{for}\ k > k^{\prime}, \\
-(k/k^{\prime})^{\ell} & \mbox{for}\ k < k^{\prime}.
\end{array} \right. 
\label{integral_I}
\end{equation}
For $\ell < 0$, since $I_{\ell}(k,k^{\prime}) > 0$, 
the potential in (\ref{pairing_potential}) acts as repulsive one and is 
inappropriate for the formation of pairing.
Hence we consider only the $\ell>0$ case.

For the singlet pairing case, 
diagonalization of the gap equation is complicated
because of the existence of tunneling term. 
In the absence of tunneling, the gap equation has block diagonalized form
and is given by the same equation (\ref{block_gap}) for triplet pairing case.
At $T=0$, the gap equation becomes
$\Delta_{\bf k} = - (\phi_2/2) P_{\bf k}$ where
$P_{\bf k}$ is given by (\ref{eq_P}) with
the replacement of $E^{\pm}_k$ by $E_k = \sqrt{\xi_k^2 + |\Delta_k|^2}$.
An explicit calculation shows that $I_{\ell=0}(k,k^{\prime})=0$, which implies
that the singlet {\it s}-wave pairing state never occurs at $\nu=1/m$ 
in double-layer quantum Hall systems.

Now, to solve the gap equation for both the singlet
and the triplet pairing case,
we introduce an approximation for $|\Delta_k|$.
Substituting (\ref{integral_I}) into the expression of 
$P^{\pm}_{\bf k}$ and $P_{\bf k}$,
we see that $|P^{\pm}_{\bf k}|, |P_{\bf k}|= O(k^{\ell})$ 
for $k \rightarrow 0$ and
$|P^{\pm}_{\bf k}|, |P_{\bf k}|= O(k^{-\ell})$ for $k \rightarrow \infty$.
Since $\Delta_{\bf k}$ is linear in $P^{\pm}_{\bf k}$ or $P_{\bf k}$,
$\Delta_{\bf k}$ has the same asymptotic nature as 
$P^{\pm}_{\bf k}$ and $P_{\bf k}$.
This consideration leads us to employ the approximation 
for $|\Delta_k|$;
\begin{equation}
|\Delta_k| = \left\{ \begin{array}{ll} 
 \Delta_{k_F} \left( k_F/k \right)^{\ell} & \mbox{for}\ k>k_F, \\
 \Delta_{k_F} \left( k/k_F \right)^{\ell} & \mbox{for}\ k<k_F. 
                 \end{array} \right.
\end{equation}

In the absence of tunneling, the gap equation is given by
\begin{equation}
\frac{1}{\phi_2} = \frac{1}{2M} \left[
\int_0^{k_F} dk \frac{k}{E_k} \left( \frac{k}{k_F} \right)^{2\ell}
+\int_{k_F}^{\infty} dk \frac{k}{E_k} \left( \frac{k_F}{k} \right)^{2\ell}
\right]
\label{gap_eq_t=0}
\end{equation}
and the energy difference between the pairing state and 
the {\it no}-pairing state is given by
\begin{equation}
\langle H^{eff}_0 \rangle_{\Delta} -  
\langle H^{eff}_0 \rangle_{\Delta=0} = \sum_k |\xi_k| 
 \left( 1- \frac{|\xi_k|}{E_k} \right) - \frac{M \Omega}{2 \pi} 
 \frac{|\Delta_{k_F}|^2}{\phi_2}
\label{gs_energy}
\end{equation}
for both singlet and triplet pairing with $(a,b)=(0,1)$.
A numerical estimation of (\ref{gs_energy}) reveals that
the pairing state {\it always} has lower energy than
the {\it no}-pairing state.
Eq. (\ref{gap_eq_t=0}) indicates that lower $\ell$ gives larger $\Delta_{k_F}$,
which results in the lower energy of the system from (\ref{gs_energy}).
Thus the pairing with the lowest value of $\ell$ 
is realized in the absence of the tunneling.
As mentioned above, {\it s}-wave($\ell=0$) pairing never occurs. 
Hence the lowest value of $\ell$ is $\ell=1$ and the triplet 
{\it p}-wave pairing is realized in the absence of the tunneling effect.

The next step is to take into account the tunneling effect.
From (\ref{gap_equation1}) and (\ref{gap_equation2}) we obtain,
\begin{eqnarray}
a  = \phi_1
 & & \left[ (a+b) F_1 ( |a+b|^2 \Delta^2, \tau ) 
       + (a+b) F_2 ( |a+b|^2 \Delta^2, \tau ) \right. \nonumber \\
 & & \left. + (a-b) F_1 ( |a-b|^2 \Delta^2, -\tau ) 
       + (a-b) F_2 ( |a-b|^2 \Delta^2, -\tau ) \right],
\label{gap_eq1d}
\end{eqnarray}
\begin{eqnarray}
b = \phi_2
 & & \left[ (a+b) F_1 ( |a+b|^2 \Delta^2, \tau ) 
       +(a+b) F_2 ( |a+b|^2 \Delta^2, \tau ) \right. \nonumber \\
 & & \left. - (a-b) F_1 ( |a-b|^2 \Delta^2, -\tau ) 
       -(a-b) F_2 ( |a-b|^2 \Delta^2, -\tau ) \right],
\label{gap_eq2d}
\end{eqnarray}
where $\Delta = \Delta_{k_F}/\epsilon_F$, $\tau = t/\epsilon_F$,
$F_1(g,\tau) = 1/2 \int_0^1 dy 
   y/ \sqrt{ (y-1-\tau)^2 + g y }$,
$F_2(g,\tau) = 1/2 \int_0^1 dy
   1/ \sqrt{((1+\tau)y-1)^2 + g y^{3} }$.
First we discuss the $\nu=1/2$ double-layer quantum Hall systems.
In this case we set $(\phi_1,\phi_2)=(2,2)$.
From (\ref{gap_eq1d}), (\ref{gap_eq2d}) and $|a|^2+|b|^2=1$,
we calculated numerically the values of $|a|$, $|b|$ and $\Delta$. 
In FIG.\ref{nu_half}, we show the $\tau$
dependence of $\Delta$ and $\theta \equiv \tan^{-1} \left( |a|/|b| \right)$.
At $\tau=0$, $(a,b)$ is equal to $(0,1)$ 
which corresponds to the $(3,3,1)$-state.
As $\tau$ approaches to $2$, $\theta$ increases up to $\pi/4$, 
and in this limit $(a,b)=(1/\sqrt{2},1/\sqrt{2})$(up to a phase factor)
which corresponds to the {\it Pfaffian}-state.
Considering the Coulomb interaction at the level of gap equation,
we find numerically that the gap decreases 
from $\Delta \sim 3.13$ to $\Delta \sim 2.49$
at $d=0$ and $\alpha \equiv (e^2/\epsilon \ell_B)/\hbar \omega_c =1$.
Since the double-layer system at the $d< \ell_B$ region can also be regarded
as the single-layer system, 
this result seems to indicate that the quantum Hall state 
is realized even at $d=0$ in double-layer systems within our treatment. 
This seems to be a contradiction with our knowledge that the quantum Hall
effect never occurs in the single layer system with $\nu=1/2$.
To treat this problem correctly, we have to include the
effects of charge density unbalance between two layers
and the Zeeman energy for real spin. 
This problem is considered in the subsequent publication \cite{morinari}.
On the other hand at $\nu=1$, we set $(\phi_1,\phi_2)=(0,2)$.
Since we can see that $(a,b)=(0,1)$ from (\ref{gap_eq1d}), 
the system has only one type of {\it p}-wave pairing state.
The gap equation becomes,
\begin{equation}
F_1(\Delta^2,\tau)+F_2(\Delta^2,\tau)
+F_1(\Delta^2,-\tau)+F_2(\Delta^2,-\tau)=1/2.
\end{equation}
In FIG.\ref{nu_1}, we show the tunneling dependence of $\Delta$.
The typical value of $\tau$ is lower than $1$ in experiments.
In this region, $\Delta$ is almost unaffected by the existence of tunneling.

In conclusion, we derive the gap equation for general pairing states
under the existence of tunneling.
Analysis of this gap equation shows that the pairing state 
is {\it p}-wave. 
At $\nu=1/2$ we observed that with increase of the tunneling amplitude
the system evolves from the $(3,3,1)$-state toward the {\it Pfaffian}-state.
At $\nu=1$ 
the state is also the CF {\it p}-wave pairing state.
To check the phase diagram obtained experimentally by Murphy et al.
\cite{murphy}
at $\nu=1$, we must describe the quantum Hall state
under the choice of $(\phi_1,\phi_2)=(2,0)$.
However, the relation between this pairing state and the quantum Hall state
is unclear and is left as future study.
Apart from the $\nu=1/2$ and $\nu=1$ case, on which we have mainly discussed
in this paper, the {\it p}-wave pairing state is also expected 
in the $\nu=1/m$ case. For instance at $\nu=1/3$, as we have mentioned earlier,
we expect two types of {\it p}-wave pairing state to emerge 
with changing the layer separation $d$.
In closing the paper, 
we remark here that our discussion can also be applied to 
the single-layer quantum Hall systems with real spin degrees of freedom 
and it will be presented in a subsequent publication \cite{morinari}.

I would like to thank N. Nagaosa and S. Murakami 
for critical reading of the manuscript.
This work is supported by JSPS.

\begin{figure}
\caption{The angle $\theta$ vs the tunneling strength $\tau$ 
at $\nu=1/2$. Inset: The gap $\Delta$ vs $\tau$.}
\label{nu_half}
\end{figure}

\begin{figure}
\caption{The gap $\Delta$ vs the tunneling strength $\tau$ 
at $\nu=1$.}
\label{nu_1}
\end{figure}


\begin{references}
\bibitem{tsui} D. C. Tsui, H. L. Stormer, and A. C. Gossard, Phys. Rev. Lett.
{\bf 48}, 1559 (1982).
\bibitem{laughlin} R. B. Laughlin, Phys. Rev. Lett. {\bf 50}, 1395 (1983).
\bibitem{jain} J. K. Jain, Phys. Rev. Lett. {\bf 63}, 199 (1989);
Phys. Rev. B {\bf 40}, 8079 (1989); {\it ibid}. {\bf 41}, 7653 (1990).
\bibitem{exp} Y. W. Suen, L. W. Engel, M. B. Santos, M. Shayegan, 
and D. C. Tsui,  Phys. Rev. Lett. {\bf 68}, 1379 (1992);
J. P. Eisenstein, G. S. Boebinger, L. N. Pfeiffer, K. W. West, and S. He,
Phys. Rev. Lett. {\bf 68}, 1383 (1992).
\bibitem{numerical} T. Chakraborty and P. Pietilainen, Phys. Rev. Lett. 
{\bf 59}, 2784 (1987); D. Yoshioka, A. H. MacDonald, and S. M. Girvin, 
Phys. Rev. B {\bf 39}, 1932 (1989); H. A. Fertig, Phys. Rev. B 
{\bf 40}, 1087 (1989).
\bibitem{halperin} B. I. Halperin, Helv. Phys. Acta. {\bf 56}, 75 (1983).
\bibitem{moore_read} G. Moore and N. Read, Nucl. Phys. B {\bf 360}, 362 (1991).
\bibitem{greiter} M. Greiter, X. G. Wen and F. Wilczek, Nucl. Phys. B 
{\bf 374}, 567 (1992).
\bibitem{ho} T. L. Ho, Phys. Rev. Lett. {\bf 75}, 1186 (1995).
\bibitem{bonesteel} N. E. Bonesteel, I. A. McDonald and C. Nayak, Phys. Rev.
Lett. {\bf 77}, 3009 (1996); 
N. E. Bonesteel, Phys. Rev. B {\bf 48}, 11484 (1993).
\bibitem{mahan} G. D. Mahan, {\it Many-Particle Physics}
(Plenum Press, New York, 1990)
chapter 10.
\bibitem{zhang} S. C. Zhang, T. H. Hansson, and S. Kivelson, 
Phys. Rev. Lett. {\bf 62}, 82 (1989); 
S. C. Zhang, Int. J. Mod. Phys. B {\bf 6}, 25 (1992).
\bibitem{boson} X. G. Wen and A. Zee, Phys. Rev. Lett. {\bf 69}, 1811 (1992);
Z. F. Ezawa and A. Iwazaki, Int. J. Mod. Phys. B {\bf 19}, 3205 (1992).
\bibitem{lopez_fradkin} A. Lopez and E. Fradkin, 
Phys. Rev. B {\bf 51}, 4347 (1995).
\bibitem{haldane_rezayi} F. D. M. Haldane and E. H. Rezayi, 
Phys. Rev. Lett. {\bf 60}, 956 (1988).
\bibitem{bcs} J. Bardeen, L. N. Cooper and J. R. Schrieffer, 
Phys. Rev. {\bf 106}, 162 (1957); {\bf 108}, 1175 (1957).
\bibitem{rajaraman} R. Rajaraman, Phys. Rev. B {\bf 56}, 6788 (1997).
\bibitem{spinless} Similar pairing interaction for spinless fermion case 
was discussed in \cite{greiter}. Their analysis helps us in the following 
discussion about gap equation.
\bibitem{bw} R. Balian and N. R. Werthamer, Phys. Rev. {\bf 151}, 1553 (1963);
P. W. Anderson and W. F. Brinkman, in {\it The Helium Liquids(Proceedings of 
the 15th Scottish Universities Summer School, 1974)}, ed. J. G. M. Armitage 
and I. E. Farquhar, pp.315-416 (Academic Press, New York, 1975).
\bibitem{meanfield} Here the gap equation is derived at the mean field level.
However, the strong coupling effect seemsto be important 
because the coupling constant is order of unity.
The strong coupling effect is not discussed here.
\bibitem{murphy} S. Q. Murphy, J. P. Eisenstein, G. S. Boebinger, 
L. N. Pfeiffer, and K. W. West,  Phys. Rev. Lett. {\bf 72}, 728 (1994).
\bibitem{morinari} T. Morinari (unpublished).
\end{references}
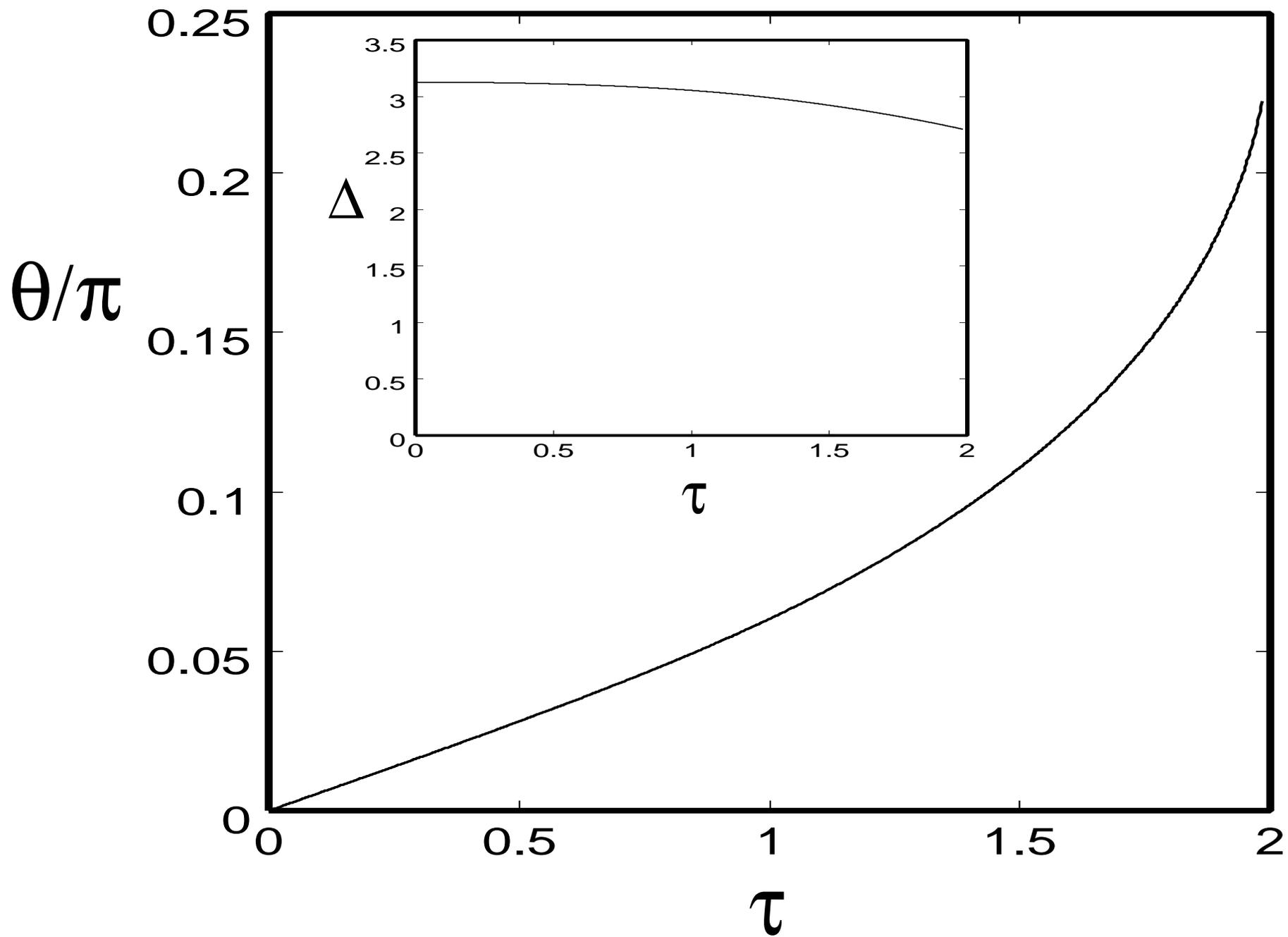
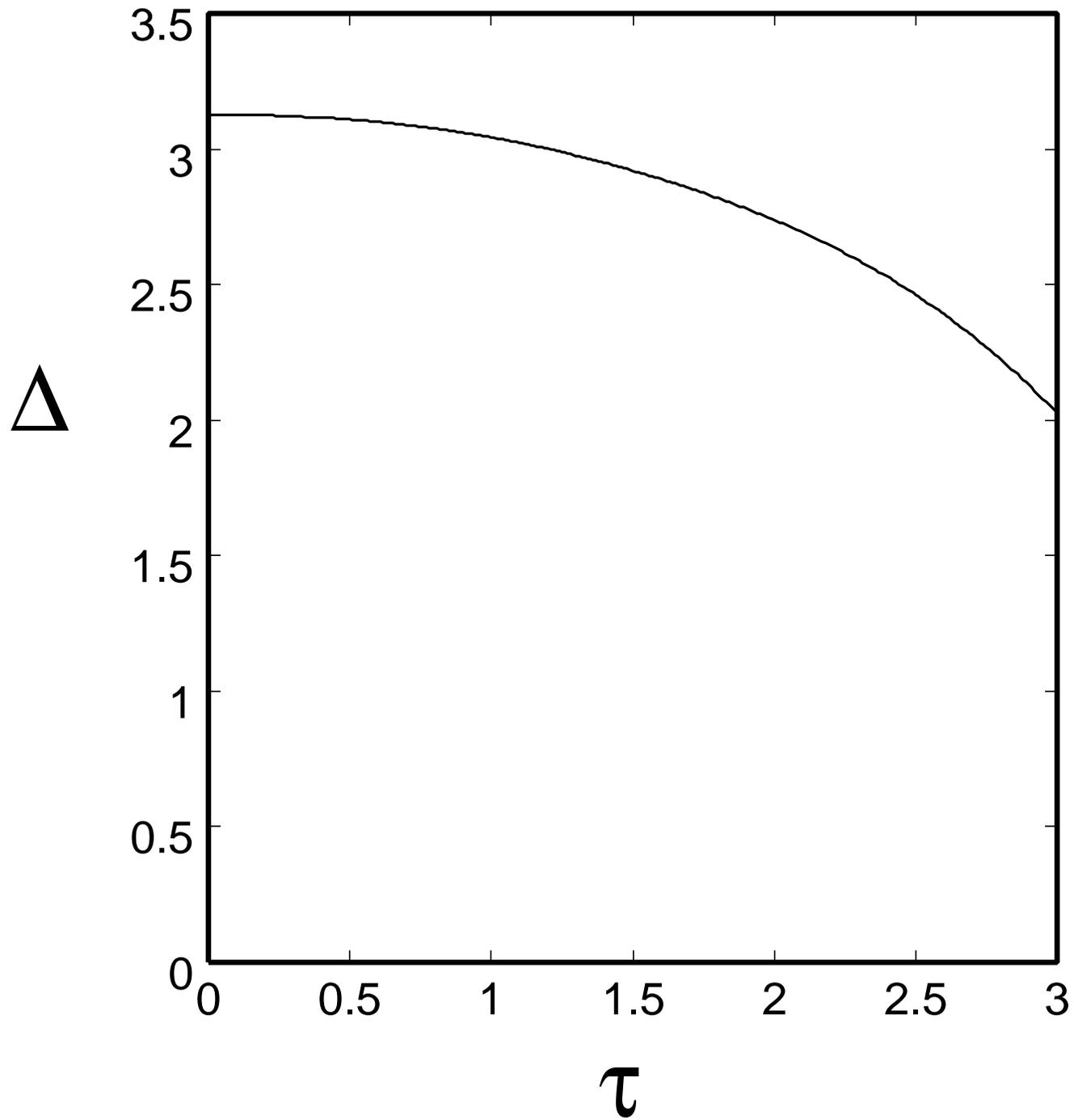
\end{document}